\documentclass[a4paper]{article}

\usepackage{INTERSPEECH2019}
\usepackage{paralist}

\title{A Transfer Learning Method for Speech Emotion Recognition from Automatic Speech Recognition}
\name{Sitong Zhou$^1$,Homayoon Beigi$^2$}
\address{
  $^1$Columbia University\\
  $^2$Recognition Technologies, Inc. and Columbia University}
\email{$^1$sz2629@columbia.edu, $^2$beigi@recotechnologies.com}

\begin{document}

\maketitle

\begin{abstract}
This paper presents a transfer learning method in speech emotion recognition based on a Time-Delay Neural Network (TDNN) architecture. A major challenge in the current speech-based emotion detection research is data scarcity. The proposed method resolves this problem by applying transfer learning techniques in order to leverage data from the automatic speech recognition (ASR) task for which ample data is available. Our experiments also show the advantage of speaker-class adaptation modeling techniques by adopting identity-vector (i-vector) based features in addition to standard Mel-Frequency Cepstral Coefficient (MFCC) features.\cite{r:beigi-sr-book-2011} We show the transfer learning models significantly outperform the other methods without pretraining on ASR. The experiments performed on the publicly available IEMOCAP dataset which provides 12 hours of emotional speech data. The transfer learning was initialized by using the Ted-Lium v.2 speech dataset providing 207 hours of audio with the corresponding transcripts. We achieve the highest significantly higher accuracy when compared to state-of-the-art, using five-fold cross validation. Using only speech, we obtain an accuracy 71.7\% for anger, excitement, sadness, and neutrality emotion content.

\end{abstract}

\section{Introduction}

Detecting emotions from speech has attracted attention for its usage in enhancing natural human-computer interaction. The ability of understanding human emotion status is helpful for machines to bring empathy in various applications. 

Speech emotion recognition is suffering from insufficiency of labeled data. Though several emotion datasets have been released \cite{r-m:busso-2008}\cite{r-m:emo-db-german-2006}\cite{r-m:savee-db-2011}, the size of emotion datasets are relatively small due to the expensive collection costs compared with plentiful data for tasks like Automatic Speech Recognition(ASR)\cite{r-m:rousseau-2014} and speaker recognition\cite{r-m:nagrani-2018}. Most speech emotion detection models are trained from scratch within a single dataset\cite{r-m:lee-2015-1}\cite{r-m:kim-2013}\cite{r-m:satt-2017}\cite{r-m:chernykh-2018}, therefore cannot successfully adapt to novel scenarios which has not been encountered during training. One possible solution is to leverage knowledge acquired from large-scale datasets of relevant speech tasks to emotion recognition domain. Although some efforts were spent on transfer learning method for categorical emotion detection from other paralinguistic tasks, such as speaker, and gender recognition and effective emotional attributes prediction\cite{r-m:latif-2018}\cite{r-m:ghosh-2016}, however they don’t  choose source domain with large-scale datasets, not shown significant improvement over non-transfer learning methods.
 
Previous research has shown that speech emotion detection can be improved after combined with textual data \cite{r-m:poria-2017}. Multi-modal methods can significantly improve the emotion detection performance by incorporating lexical features from given transcripts with the acoustic features from audios \cite{r-m:beigi-2018-1}\cite{r-m:tzirakis-2017}\cite{r-m:cho-2018}.However, in real application scenario, transcripts are often absent. Although ASR can provide transcripts in real time to emotional speech data\cite{r-m:cho-2018}, it requires large language models loaded, and is computational costly when decoding sequence. Therefore, transfer learning using ASR as the source domain might be an efficient solution in emotion detection to incorporate textual features through high level features extracted in ASR models.

Another challenge for emotion recognition is that speakers express emotions in different ways, in addition, environments can affect acoustic features. Speaker adaptation is useful to capsulate speaker and environment specific information into acoustic features. iVector\cite{r-m:dehak-2011} based adaptation has been shown fast and efficient in speech recognition\cite{r-m:peddinti-2015}. We employ i-vector based speaker adaptation in emotion detection.

This paper proposes a transfer learning method to adapt ASR models in emotion recognition domain. The model is pre-trained on Tedlium2 dataset\cite{r-m:rousseau-2014}, with over 207 hours data, and fine tuned on 12 hours of emotional speech. The model architecture is TDNN-based\cite{r-m:sugiyama-1991}\cite{r:beigi-sr-book-2011}\cite{r-m:peddinti-2015} with input as speaker adapted MFCC\cite{r:beigi-sr-book-2011} features. Our experiments show the improvements in emotion detection using transfer learning from ASR to speech emotion recognition combined with speaker adaptation. The performance is evaluated on the benchmark dataset Interactive Emotional Dyadic Motion Capture (IEMOCAP)\cite{r-m:busso-2008}, with 71.7\% unweighted accuracy  among “angry”, “happy”, “sad” and “neutral” under the 5-fold cross validation strategy. Our method significantly outperforms the state-of-art strategy\cite{r-m:satt-2017}.

\section{Related Work}

Most efficient models in prior work are based on deep learning models, which can learn high-level features from low-level acoustic features. Lee’s work\cite{r-m:lee-2015-1} showed the importance of long-range context effect, and significantly improve the RNN results over DNN model. This work had been staying state-of-art for years with 63.89\% unweighted accuracy(UA), until surpassed by a hybrid approach of convolution layers and LSTM  convolutional LSTM\cite{r-m:satt-2017} with 68.8\% accuracy. Previous literature has seldomly discussed about TDNN architecture in emotion recognition. TDNN can efficiency capture temporal information as RNN and LSTM do, but is faster for its parallelization ability and lower computation costs during training\cite{r-m:peddinti-2015}, which is a desired property when training on large-scale ASR data. 

 Many approaches\cite{r-m:lee-2015-1} \cite{r-m:chernykh-2018}are speaker independent where features are normalized for individuals resulting in information loss, while our work is conducted in speaker dependent context using full MFCC raw features combined with iVectors containing speaker characteristics\cite{r-m:dehak-2011}. Peddinti\cite{r-m:peddinti-2015} has proposed an efficient TDNN-based architecture for ASR with features as the combination of MFCC and iVectors, efficiently learned robust representations among various speakers and environments. Our study uses bottleneck layers of this TDNN architecture to use high-level feature representations that reflect insights from ASR tasks, and fine tunes on emotion datasets. 
\section{Method}
\subsection{Problem Definition}
The emotion recognition problem is a classification problem when we represent emotion as categories rather than dimensional representations,  \begin{equation}
D =  \left \{  \left ( X,z \right ) \right \}
  \label{eq1}
\end{equation}
where where \(X\)  are the acoustic features input and \(z\) is dimensional output corresponding to the emotion prediction. We want to find a function
\begin{equation}
D =  \left \{ f : X \rightarrow z
 \right \}
  \label{eq1}
\end{equation} to map features to categories. This model is trained on frame-level labels, and predicts the utterance labels  by aggregating frame-level predictions through max likelihood by summing up the results of frames.

\subsection{Feature}

Full MFCC features with all 40 coefficients are computed at each time index is used as input to neural network. Instead of mean normalization on MFCC, an 100-dimension iVector is appended to MFCC features at each frame to encode mean-offset information. The iVector extraction model is trained as described in \cite{r-m:peddinti-2015}. 

\subsection{TDNN for ASR}
TDNN is designed for capturing long term temporal dependencies in lower computational costs compared to RNN. It operates similar to a feed forward DNN architecture where lower layers focus input content in narrow windows, and higher layers connects windows of selected previous layer nodes to process the information from a wider context. Therefore its deeper layers can learn effective long term temporal dependencies without recurrent connections which hurdle parallel computation. 

The pretraining on ASR follows the Kaldi recipe for the TED-Lium tasksi\cite{tedlium-recipe-r2}, where uses 13 TDNN layers and each layer consists of 1024 activation nodes. The time stride of each layer, which defines the window at which calculating over nodes at neighbor time steps in the past layer, is assigned as 0 for the 1\textsuperscript{st} and 5\textsuperscript{th} TDNN layer, as 1 from the 2\textsuperscript{nd} and 4\textsuperscript{th} layer, and as 3 for layers after since the 6\textsuperscript{th}.  A fully connected prefinal layer of 1024 dimension follows the 13\textsuperscript{th} TDNN layer before decoding output sequences. The model is trained with a sequence-level objective function named lattice-free version of the maximum mutual information (LF-MMI)\cite{r-m:povey-2016}, for maximising the log-likelihood of the correct sequences.

\subsection{Training on Emotion Labels}
Emotion labels are given for each utterance in the dataset. We label all the frames using the utterance label where the frames lie in. To train for emotion detection, the 12\textsuperscript{th} and 13\textsuperscript{th} TDNN layers as well as the ASR prefinal layer are selected to produce bottleneck embeddings as high level features learnt from ASR, a new fully connected dense layer is appended after the embedding layer for predicting the frame-level emotion labels, and a softmax layer with four dimension outputs is used to predict frame-level emotion. The model uses cross-entropy as the objective function for frame-level classification.

The output of this model is for frame unit rather than for utterance unit, we aggregate frame-level predictions,    using maximum likelihood by adding the output vectors over frames, corresponding to the highest valued dimension of the sum of the softmax layer output over all frames within the utterance

\section{Experiment}
\subsection{Dataset}
This work uses IEMOCAP, which contains 12 hours of audio data with scripted and improvised speech, performed by ten actors, one male and one female as a pair in five sessions. In training and testing, four categories “angry”, “happy”, “sad” and “neutral” are selected out of ten categories, for a more balanced and efficient dataset, resulting in a final collection of 4936 utterances, each utterance has unique emotion label. This dataset consists of five sessions, and the category distribution is as in Table~\ref{tab:emotion_distribution}.

\begin{table}[th]
  \caption{Emotion category distribution in IEMOCAP }
  \label{tab:emotion_distribution}
  \centering
  \begin{tabular}{ r|rrrr|r }
    \toprule
    \multicolumn{1}{r}{\textbf{session}} & \multicolumn{1}{r}{\textbf{ang}} &\multicolumn{1}{r}{\textbf{exc}} &\multicolumn{1}{r}{\textbf{neu}} &\multicolumn{1}{r}{\textbf{sad}} &
                                 \multicolumn{1}{r}{\textbf{total}} \\
    \midrule
    ses1   & $229$  & $143$ & $384$ & $194$ & $950$             \\
    ses2   & $137$  & $210$ & $362$ & $197$ & $906$             \\
    ses3   & $240$  & $151$ & $320$ & $305$ & $1016$              \\
    ses4   & $327$  & $238$ & $258$ & $143$ & $966$              \\
    ses5   & $170$  & $299$ & $384$ & $245$ & $1098$             \\
    \hline
    total   & $1103$  & $1041$  &  $1708$ & $1084$ & $4936$             \\

    \bottomrule
  \end{tabular}
  
\end{table}
\subsection{Pretraining on ASR}
For pretraining on ASR, we use the feed-forward TDNN to capture long term temporal dependencies from short term feature representations. Hidden activations are sub-sampled in order to speed up the training\cite{r-m:peddinti-2015}. The model is pre-trained on ASR data, with 13 TDNN layers and output layers for decoding sequence based on acoustic models. As neighboring activations shares largely overlapped  input contexts, sub-sampling on activations can reduce computational costs without sacrificing the coverage range over input frames. The hyper parameters for model architecture and training are chosen according to Kaldi Tedlium2 TDNN recipe\cite{tedlium-recipe-r2}, which has been tuned properly on Tedlium2 dataset, achieving 7.6 word error rate(WER) on test dataset after six epochs training. The parameters are optimized through preconditioned stochastic gradient descent (SGD) updates, following the training recipe detailed in \cite{r-m:zhang-2014}.

\subsection{Training for Emotion Classification}
After obtaining the pretraining model, we use 12\textsuperscript{th}, 13\textsuperscript{th} and the prefinal layer as the bottleneck features for the appended fully connected layer and the softmax layer for predicting frame emotions. We test on session 5 after training on session 1-4, and find the 12\textsuperscript{th} TDNN layer the best performance, therefore we use the 12\textsuperscript{th}layer output as the bottleneck embeddings for later experiments. 

\subsection{Evaluation Method}
For parallel comparison with other methods\cite{r-m:lee-2015-1}\cite{r-m:satt-2017}\cite{r-m:chernykh-2018}, we train under a 5-fold cross validation strategy where each time we choose a session from IEMOCAP for testing, and the other four for training. The results are evaluated by the average of unweighted accuracy over five cross validation experiment runs. 

\section{Results}
\subsection{Bottleneck Layer Selection}

Compare the prefinal, 12\textsuperscript{th} and 13\textsuperscript{th} layer as in Table~\ref{tab:bottleneck_compare}, we found that the 12\textsuperscript{th} layer has the best performance and prefinal have the worst. We hypothesis that is because in ASR, prefinal and 13\textsuperscript{th} layer are more specialized in speech recognition as they are closer to the final output layer, while the 12\textsuperscript{th} learns the general high-level acoustic features that helps emotion recognition.

\begin{table}[th]
  \caption{Test Accuracy on Session 5 with different bottleneck layers}
  \label{tab:bottleneck_compare}
  \centering
\begin{tabular}{rr}
\toprule
    \multicolumn{1}{r}{\textbf{Bottleneck Layer}} & \multicolumn{1}{r}{\textbf{Test Accuracy on Ses 5 (\%)}}\\
\midrule
Prefinal   & $63.4$    \\
13\textsuperscript{th} TDNN    & $66.7$    \\
12\textsuperscript{th} TDNN     & $69.3$  \\
\bottomrule
\end{tabular}
\end{table}

\subsection{Model performance}
Our model using the 12\textsuperscript{th} TDNN layer outperforms other current state-of-art methods. Table~\ref{tab:model_compare} to our best knowledge. The 5-fold cross validation unweighted accuracy is improved from 68.8\%\cite{r-m:satt-2017} to 71.7\%.

\begin{table}[th]
  \caption{Model Comparison on Unweighted Accuracy(UA) in \% from Five-fold Cross Validation}
  \label{tab:model_compare}
  \centering
\begin{tabular}{rrll}
\toprule
    \multicolumn{1}{r}{\textbf{Model}} & 
   \multicolumn{2}{r}{\textbf{Single Session UA}} &
    \multicolumn{1}{l}{\textbf{5-fold CV UA}}\\
\midrule
ASR Transfer Learning & ses1  & $65.3$ & {}\\
{} & ses2 & $78.9$ & {} \\
{} & ses3  & $71.1$  & {}\\
{} & ses4  & $73.9$ & {} \\
{} & ses5  & $69.3$& {}\\

{} &  {} & {} & $\textbf{71.7}$\\
\hline
RNN-ELM\cite{r-m:lee-2015-1} & {} & {} & $63.9$   \\
Conv-LSTM\cite{r-m:satt-2017} & {} & {} & $68.8$ \\
CTC-BLSTM\cite{r-m:chernykh-2018} &  {} & {} & $54$\\
\bottomrule
\end{tabular}
\end{table}

\subsection{Error analysis}
\begin{figure}[t]
  \centering
  \includegraphics[width=\linewidth]{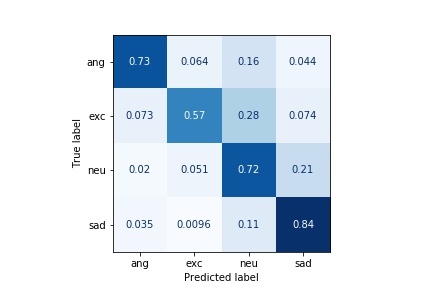}
  \caption{Confusion Matrix for Transfer Learning Method Predictions}
  \label{fig:confusion_matrix}
\end{figure}

To study the model performance within each emotion category, we present a confusion matrix,  shown in  Figure~\ref{fig:confusion_matrix}, by calculating the average confusion matrix of five cross validation experiments for our best model architecture, using TDNN 12\textsuperscript{th} layer as the bottleneck layer. The excitement category has a lower accuracy compared to other categories, in which 28\% samples are confused with neutral utterances. We observed that the netral emotion is the most likely wrong prediction from all non-neutral categories. It might due to the fact that non-neutral utterances usually consist of a large proportion of frames carrying no emotional content. We also found the model confuses neutrality as sadness, that may because many neutral utterances are in low tone, as in sadness.

\section{Conclusion}

Our study shows transfer learning from ASR is a good strategy for emotion classification, and indicates potential feature overlap between speech-to-text and emotion recognition. Our method is limited in frame-level prediction, where frames are predicted first then aggregated into utterance level labels. The frame-level structure results in the ignorance of sequential information in emotion labels decoding. In future, we expect sequence models can predict at utterance-level and bring further performance improvements by considering sequential information for sequence decoding. 

\bibliographystyle{IEEEtran}
\bibliography{ms}
\end{document}